\documentclass[superscriptaddress,
 amsmath,amssymb,
 aps,
]{revtex4-2}

\usepackage{graphicx}
\usepackage{bm}

\begin{document}

\preprint{APS/123-QED}

\title{Strong Quantum Turbulence in Bose Einstein Condensates}

\author{H. A. J. Middleton-Spencer}
\email{h.a.j.middleton-spencer@bham.ac.uk}
\affiliation{%
Joint Quantum Centre Durham-Newcastle, School of Mathematics, Statistics and Physics, Newcastle University,
Newcastle upon Tyne NE1 7RU, United Kingdom
}%
\affiliation{%
School of Physics and Astronomy, University of Birmingham, Edgbaston, Birmingham B15 2TT, United Kingdom
}%
\author{A. D. G. Orozco}%
\affiliation{%
Instituto de Física de S\~{a}o Carlos, Universidade de S\~{a}o Paulo, Av. Trabalhador S\~{a}o-carlense, 400 Pq.
Arnold Schimidt, 13566-590 S\~{a}o Carlos, SP, Brazil
}
\author{L. Galantucci}%
\affiliation{%
Joint Quantum Centre Durham-Newcastle, School of Mathematics, Statistics and Physics, Newcastle University,
Newcastle upon Tyne NE1 7RU, United Kingdom
}%
\affiliation{Istituto per le Applicazioni del Calcolo ‘M. Picone’, IAC-CNR, Via dei Taurini 19, 00185 Roma, Italy}
\author{M. Moreno}%
\affiliation{%
Instituto de Física de S\~{a}o Carlos, Universidade de S\~{a}o Paulo, Av. Trabalhador S\~{a}o-carlense, 400 Pq.
Arnold Schimidt, 13566-590 S\~{a}o Carlos, SP, Brazil
}
\author{N. G. Parker}%
\affiliation{%
Joint Quantum Centre Durham-Newcastle, School of Mathematics, Statistics and Physics, Newcastle University,
Newcastle upon Tyne NE1 7RU, United Kingdom
}%

\author{L. A. Machado}%
\affiliation{%
Instituto de Física de S\~{a}o Carlos, Universidade de S\~{a}o Paulo, Av. Trabalhador S\~{a}o-carlense, 400 Pq.
Arnold Schimidt, 13566-590 S\~{a}o Carlos, SP, Brazil
}
\author{V. S. Bagnato}%
\affiliation{%
Instituto de Física de S\~{a}o Carlos, Universidade de S\~{a}o Paulo, Av. Trabalhador S\~{a}o-carlense, 400 Pq.
Arnold Schimidt, 13566-590 S\~{a}o Carlos, SP, Brazil
}
\author{C. F. Barenghi}%
\affiliation{%
Joint Quantum Centre Durham-Newcastle, School of Mathematics, Statistics and Physics, Newcastle University,
Newcastle upon Tyne NE1 7RU, United Kingdom
}%

\date{\today}

\begin{abstract}
By combining experiments and numerical simulations which model the dynamics of shaken atomic Bose-Einstein condensates, we reveal the surprising nature of quantum turbulence in these systems. Unlike the tangles of vortex lines described in the superfluid helium literature, we find that {our} turbulent atomic condensate contains a mixture of strong fragmented density fluctuations and small random vortex loops which are not homogeneously distributed. This unusual form of turbulence, with its own properties and scaling behaviour, which we call strong quantum turbulence, is significantly different from the turbulence which is observed in either classical or other quantum systems, thus posing a new challenge in turbulence research.
\end{abstract}

\maketitle

\noindent
Quantum fluids (e.g. superfluid helium, atomic Bose-Einstein condensates,
polariton condensates and the interior or neutron stars) differ from ordinary fluids in two respects. The first is superfluidity (the absence of viscosity), a consequence of the particular dispersion relation of elementary excitations. The second is more fundamental: the quantisation of the circulation, a direct consequence of the existence of a macroscopic wavefunction describing the properties of the whole atomic cloud. This second property implies that vorticity is constrained to vortex lines of fixed strength, proportional to Planck's constant. Quantum turbulence (turbulence of a quantum fluid) thus consists of a disordered tangle of vortex lines moving in an inviscid background \cite{BarenghiSkrbekSreeni}, unlike classical turbulence (turbulence of ordinary viscous fluids such as air or water) \cite{Frisch1995} where vortices are unconstrained in size and strength and are diffused by viscosity.

\bigskip

Most work on quantum turbulence has been carried out in superfluid helium (both $^4$He and $^3$He), the driving concern being the comparison with classical turbulence. Indeed, dedicated cryogenic techniques to visualize vortex lines \cite{LaMantia2014,Guo2010} and measure velocity fields \cite{maurer1998local,duri2015,salort2011} have led to the discovery of remarkable similarities between quantum turbulence and classical turbulence \cite{BarenghiLvovRoche}. A major observed similarity   \cite{maurer1998local,Salort2010turbulent} is the $k^{-5/3}$ Kolmogorov \cite{kolmogorov-1941} energy spectrum (where $k$ is the wavenumber) describing the distribution of kinetic energy over the length scales; this property is the signature of a universal effect: an energy cascade from large to short length scales. Further experiments have also uncovered non-classical aspects of quantum turbulence  \cite{walmsley2008quantum,BarenghiSergeevBaggaley,lamantia-duda-rotter-skrbek-2013,galantucci-sciacca-2014} which still need to be understood.
\bigskip

Unlike liquid helium, the properties of atomic condensates are
exquisitely tunable, opening the possibility of studying the phenomenon of turbulence (still a major challenge for scientists) in a more general way. Unfortunately, the study of three-dimensional (3D) quantum turbulence in atomic condensates has been frustrated by difficulties in directly visualizing vortices and in measuring turbulent velocity fields (necessary to quantify intensity and structure of the turbulence); it is worth noticing that the absence of these difficulties in 2D condensates has allowed great progress in 2D quantum turbulence \cite{reeves2013inverse,gauthier-etal-2019,johnstone-etal-2019}).
Nevertheless, pioneering work with 3D condensates confined by harmonic traps \cite{thompson-etal-2013} or boxtraps \cite{navon2016emergence} has shown evidence of energy transfer from large to small length scales. However, attempts to observe the same scaling behaviour measured in turbulent superfluid helium and in classical turbulence have not been successful, partly also due to the limited $k$-space available in atomic condensates given their small size.
\bigskip

Our work tackles these difficulties. By combining experiments and numerical simulations which model the excitation of the condensate by shaking the confining external potential, we reveal that, surprisingly, turbulence in {our} atomic condensate is qualitatively different from the familiar tangles of vortex lines described in the superfluid helium literature and in previous numerical models of turbulent condensates \cite{PhysRevA.76.045603,PhysRevLett.104.075301}, which in hindsight appear idealized. Instead, we find that the condensate contains huge nonlinear density waves, almost fragmenting the cloud. The vortex lines take the form of very short vortex loops, randomly oriented but distributed non-homogeneously, instead of the familiar distribution of long and short vortex lines observed in experiments and numerical simulations of turbulent helium. Such mixture of strong nonlinear waves and small random vortex loops non-homogeneously distributed has never been reported in a turbulent system, either quantum or classical, and presents a new challenge in turbulent research.
\bigskip
\newline
{\bf Exciting the condensate.}
Various techniques to excite turbulence in condensates have been proposed, such as phase imprinting singly-charged \cite{PhysRevLett.104.075301} or
multicharged \cite{PhysRevA.96.023617} vortices, or rotating the condensate \cite{PhysRevLett.95.145301,PhysRevA.76.045603}. Here we focus on the successful technique of shaking the trap confining the condensate \cite{PhysRevLett.103.045301,navon2016emergence}. 
This technique was first shown to create vortices \cite{PhysRevA.79.043618,Seman2011RouteTT,PhysRevA.82.033616}, before being further developed in a number of notable studies revealing in particular the presence of a turbulent cascade \cite{GarciaOrozco2020}.
\bigskip

{In the experiment, we prepare a $^{87}$Rb condensate of $N=3\times 10^5$ atoms within a cigar-shaped harmonic trapping potential (more details about experimental techniques in Methods and in Refs. \cite{PhysRevA.79.043618,PhysRevLett.103.045301,Seman2011RouteTT}). The condensate is driven for a time $T_D$ using an secondary oscillatory magnetic trap, followed by a waiting time $T_H$, in which the condensate is kept in the static harmonic trap, before finally released and imaged. In experiments, images of the condensate are typically taken from a light source which has travelled through the expanding condensate in the imaging plane after the trapping harmonic potential has been switched off. During the time-of-flight (TOF), the momentum distribution is obtained by observing the number of atoms travelling different distances from the beginning of the ballistic motion. Our group \cite{bahrami2015investigation,PhysRevA.106.023314} and others \cite{navon2016emergence} have demonstrated the validity of the TOF technique to obtain the 2D column-integrated momentum distribution ${n}(k)$ for a self-similar turbulent cloud.}

\bigskip

{\textbf{Computational methods}} The experiment is simulated using the Gross-Pitaevskii equation (see Methods for details) non-dimensionalised via harmonic oscillator units as follows

\begin{equation}
    i\frac{\partial\Psi}{\partial t}=-\frac{1}{2}\nabla^2 \Psi+\mathcal{C} |\Psi|^2\Psi+V\Psi-\mu\Psi.
    \label{eq:gpe_hou}
\end{equation}
The solution of Eq. (\ref{eq:gpe_hou}) depends on {two} dimensionless parameters. The first parameter, $\mu$, is the chemical potential, which dictates the size of the condensate. The second parameter, $\omega_z$, appears in the axisymmetric trapping potential
 
\begin{equation}
    V(x,y,z) = \frac{1}{2}\Big[(x^2+y^2)+\omega_z^2z^2\Big],
    \label{eq:harmonic_potential}
\end{equation}
and sets the geometry (oblate, spherical or prolate) of the condensate.  
{The parameter $\mathcal{C}$,  dependent on $\mu$ and $\omega_z$, denotes the interaction strength (which in our case is positive, signifying repulsive interactions). For computational feasibility, we increase the experimental value of $\omega_z=0.16$ to $0.5$ and lower the chemical potential {from $\mu=13$} to $\mu=8$. The parameter $\mathcal{C}$ is set to $1715$ (see Methods).}
\bigskip

To generate turbulence  we shake the condensate by superimposing an oscillatory potential of the form
\begin{equation}
    V_{\text{osc}}(x,y,z,t) = A\mu[1-\cos{(\Omega t})]z'/R_z,
    \label{eq:driving_potential}
\end{equation}
\noindent
to the harmonic trapping potential $V$, in Eq.~(\ref{eq:harmonic_potential}), where $A$, $\Omega$ and {$z'/R_z$} denote respectively the amplitude, the frequency and the length of the driving, $R_z$ being the Thomas-Fermi radius in the $z$ direction. We match the value of the amplitude $A$, frequency $\Omega$, and time $T_D$ to the experiment with $A=1.25$, and $\Omega=0.97$ and $T_D=10\pi/\Omega=32.4$ (all values are reported in non-dimensional units). The direction of the driving is $z'=\cos{(\theta_z)}z -\sin{(\theta_z)}x$ where $\theta_z=5^\circ$ breaks the symmetry of the system around the $z$-axis \cite{PhysRevA.79.043618,PhysRevA.79.043618,PhysRevLett.103.045301,Seman2011RouteTT,shiozaki2011transition}. For $t>T_D$, the condensate is left to evolve and oscillate for $T_H$ in the static harmonic potential, Eq. (\ref{eq:harmonic_potential}). 
\bigskip

\begin{figure*}
\centering
{\includegraphics[trim={2cm 1cm 0.5cm 1cm},clip,width=0.99\textwidth]{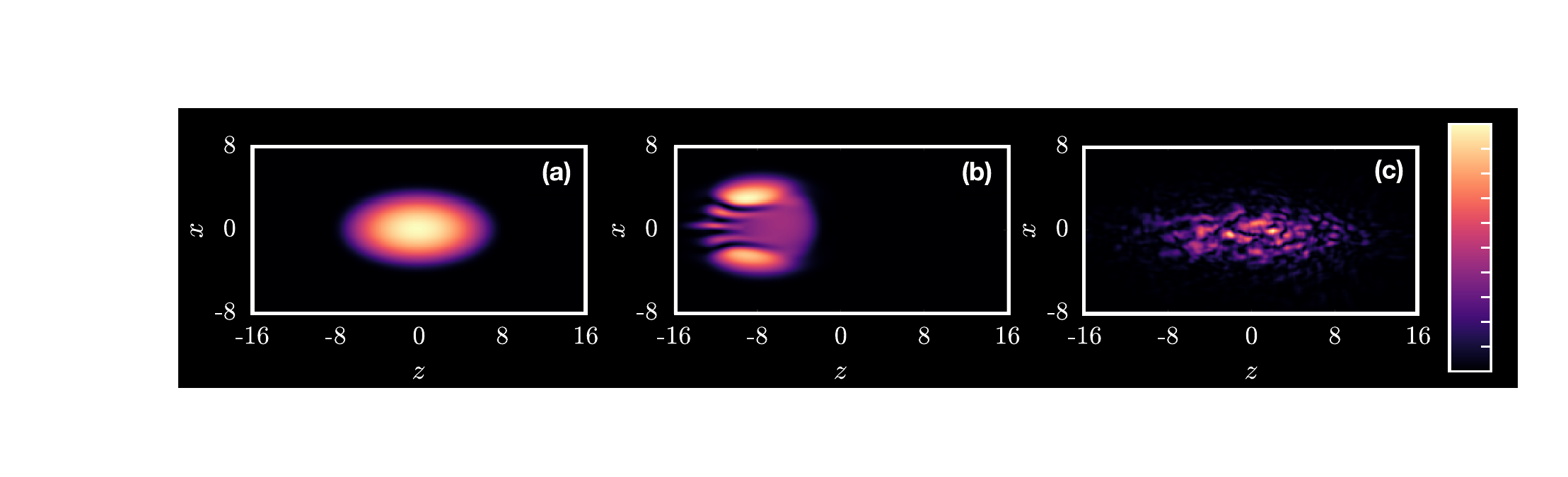}}
\caption{2D density slices 
$n(x,y=0,z)$ of the simulated condensate at (a)
$t=0.0$ (the ground state), (b) $t=4.9$ (nucleation of solitons) and (c) $t=35.2$ (turbulent state with vortices and strong density waves). 
}
\label{fig:Fig1}
\end{figure*}

{\bf Onset of turbulence.} Figure~\ref{fig:Fig1} shows the shape of the condensate during the evolution, from (a) the initial ground state, to (b) the generation of deep density waves in the form of dark solitons, to (c) the turbulent state.

\begin{figure*}[ht]
    \centering
{\includegraphics[trim={0cm 0.0cm 0.95cm 0.8cm},clip,width=0.95\textwidth]{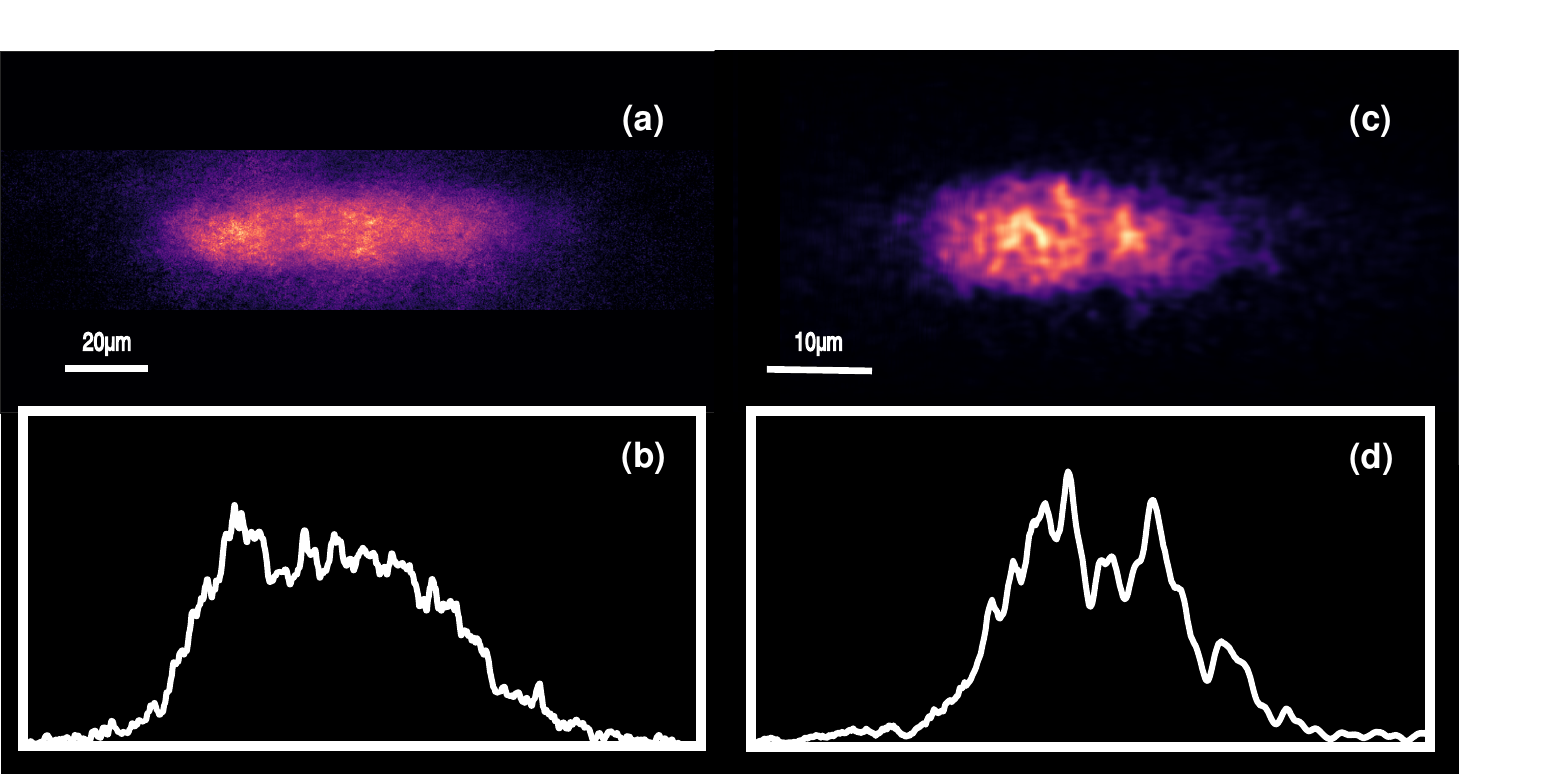}}
    \caption{{Panel (a) shows an {experimental} absorption image of the turbulent condensate taken after a 30ms TOF. Panel (c) shows the expanding computed 2D column-integrated density $n(x,z)$ (see Methods for details in the numerical expanded density). Panels (b) and (d) {respectively} correspond to {experimental and numerical} 1D slices of the integrated densities at $x=0$, showing large fluctuations on top of a background density}. {Units of length, in SI form, are added to each TOF image for reference.}}
    \label{fig:Fig2}
\end{figure*}

It is natural to ask what is nucleation process of vortex lines in the absence of an external, small-scale stirring potential (`laser spoon') \cite{Desbuquoise,kwon-etal-2021}. The process is more clear during the first oscillation of the condensate, before it is masked by large density fluctuations. For a large driving amplitude ($A>1$), dark solitons 
(nonlinear waves characterised by a localised dip in the density and a step in the condensate phase)
appear at the front of the condensate moving in the $-z$ direction (Fig. \ref{fig:Fig1}(b)). Solitons have previously been generated in condensates via a variety of techniques but are stable only in quasi-1D systems \cite{frantzeskakis2010dark}. Indeed, the solitons quickly break down into vortex lines and sound waves \cite{Mossman}, which respectively multiply and grow in size as the shaking continues (Fig. \ref{fig:Fig1}(c)).

\bigskip
For $t>T_D$, the condensate moves unforced and undergoes large oscillations about the minimum of the trapping potential, with a detectable breathing mode also present. During the oscillation, the condensate retains a shape similar to the initial cigar-shaped profile only when the centre of mass is near the the potential minimum at $z=0$; it is most distorted when far from this  minimum, near the points where the centre of mass reverses its direction. The most notable features of the obtained turbulent state at $t>T_D$ are the observed large density oscillations shown in Fig.~\ref{fig:Fig2}, where experimental absorption images, necessarily 2D  (Fig. \ref{fig:Fig2}(a)), are compared (Fig. \ref{fig:Fig2}(c)) to computed 2D column-integrated density fields $n(x,z)$, defined as
\begin{equation}
    n(x,z) = \int |\Psi(x,y,z)|^2dy.
\label{eq:n2D}
\end{equation}
To better appreciate the non-homogeneous density characterising the turbulent state, in Fig. \ref{fig:Fig2}(b), (d) we also report the corresponding 1D density profiles.
\bigskip

\begin{figure*}
    \centering
{\includegraphics[trim={0.0cm 0cm 0.0cm 0.0cm},clip,width=0.8\textwidth]{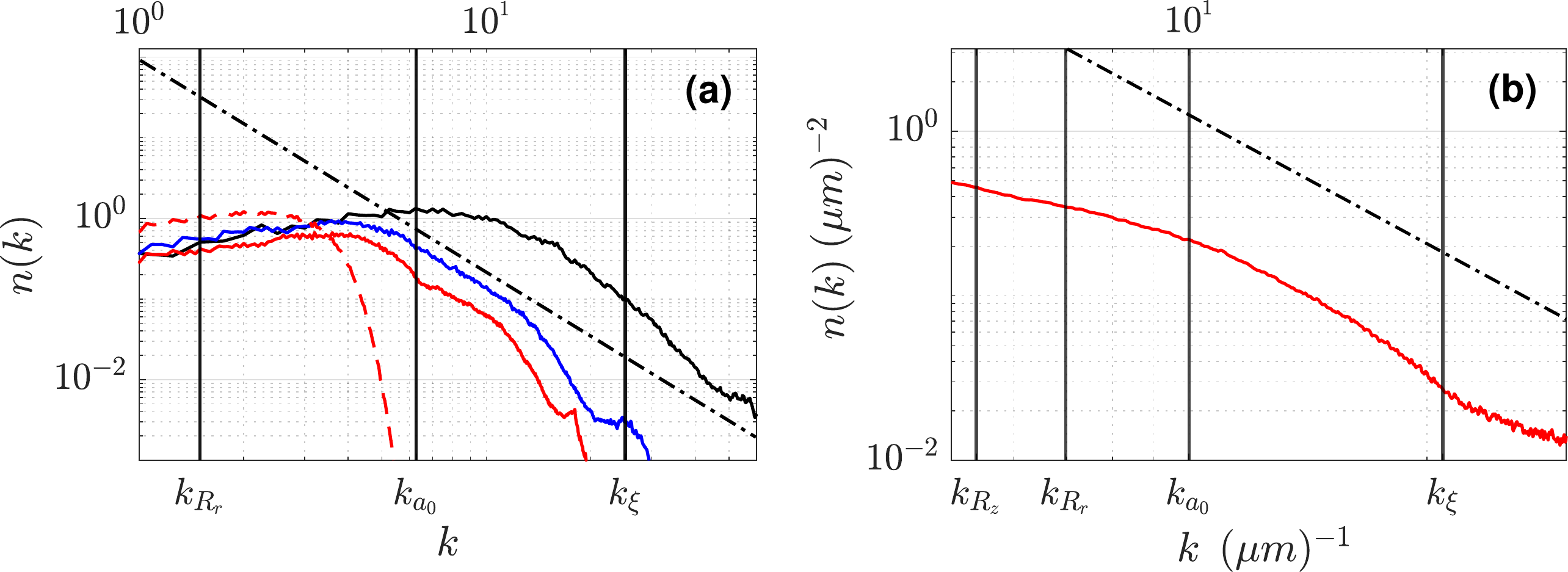}}
    \caption{The momentum distribution obtained from (a) an expanding numerical condensate {released at $t = 66$} and (b) an average of 10 experimental images of the expanding turbulent state {at an expansion time of 30ms}. The different distributions shown in (a) correspond to the expanding condensate at {expansion times} $t_{exp}=0.5$ ($t=66.5$) (black), $t_{exp}=1.5$ ($t=67.5$) (blue) and $t_{exp}=2.5$ ($t=68.5$) (red). For reference, the corresponding ground state expansion at $t_{exp}=2.5$ is given (red dashed). {We show the early time expansion of $t_{exp}=0.5$ to highlight the convergence in the later time results}. {The power law fits, superimposed to each graph for reference, are {$k^{-2.6}$} for both (a) and (b)}. Figure (b) is presented in SI units.}
    \label{fig:Fig3}
\end{figure*}

{\bf Momentum Distribution.}
To characterise the turbulence we examine
the momentum distribution ${n}(k)$ 
\cite{bahrami2015investigation}
where $k$ is the magnitude of the wavenumber. Of particular interest here is the comparison with the momentum distribution
obtained from experimental 2D density absorption images. For this
purpose, we need to account for the expansion of the 
condensate. This is done by numerically simulating the expansion  
in the frame of reference of the centre of mass after the 
trapping potential is set to zero. We compute the column integrated 
2D density at different times $t_{exp}$ after the beginning of the expansion of the turbulent BEC at $t=66$, 
and, assuming ballistic expansion,
we relate the position $\mathbf{x}$ on the enlarged condensate \cite{thompson2013evidence}
to the wavenumber $\mathbf{k}$ of the atoms before the expansion; the final step consists of
computing the momentum distribution ${n}(k)$. {For simplicity, and following the experimental procedure, we assume isotropy such that $k^2=k_x^2+k_z^2$}. {Although the time of flight expansion of the simulation is much smaller than that of the experiment, the spectra quickly converges to a power law in the $k$ sub-range of $k_{4\xi}$ to $k_{2\xi}$ (see Fig. \ref{fig:alpha} in Methods for the fittings of the exponents). The exponent of the power law quickly converges; taking the average of the results after $t=68.0$ $(t_{exp}=2.0)$ (see Fig. \ref{fig:alpha}), we obtain ${n}(k) \sim k^{-2.6\pm0.1}$
in the aforementioned range between $k_{a_0} = 2\pi/a_0$ (corresponding to the vortex core 
size $a_0\approx 4\xi$) and $k_{2\xi}=2\pi/2\xi$ (where $\xi$ is
the healing length)}. This distribution,
shown in Fig.~\ref{fig:Fig3}(a),
compares well with the experimental distribution ${n}(k) \sim k^{-2.60}$
(obtained by averaging 10 experimental runs) shown in
Fig.~\ref{fig:Fig3}(b). 
This good agreement confirms
the accurate modeling of the experiment, 
but it must be stressed that
the  range of wavenumbers where the scaling takes place is narrow.
This limitation arises from the small size of typical atomic condensates, and
prevents good quantitative comparison with turbulent superfluid helium
and classical turbulence. In Fig. \ref{fig:Fig3} we also report the momentum distribution of the expanded ground state which clearly shows how the excitation of the condensate has triggered an energy transfer towards small scales.

In the next section, we will show
that large amplitude sound-waves, fragmentation and small vortex lines and loops are responsible for this incompressible energy transfer.

\begin{figure*}[]
\centering
{\includegraphics[trim={1cm 13cm 1cm 9cm},clip,width=0.95\textwidth]{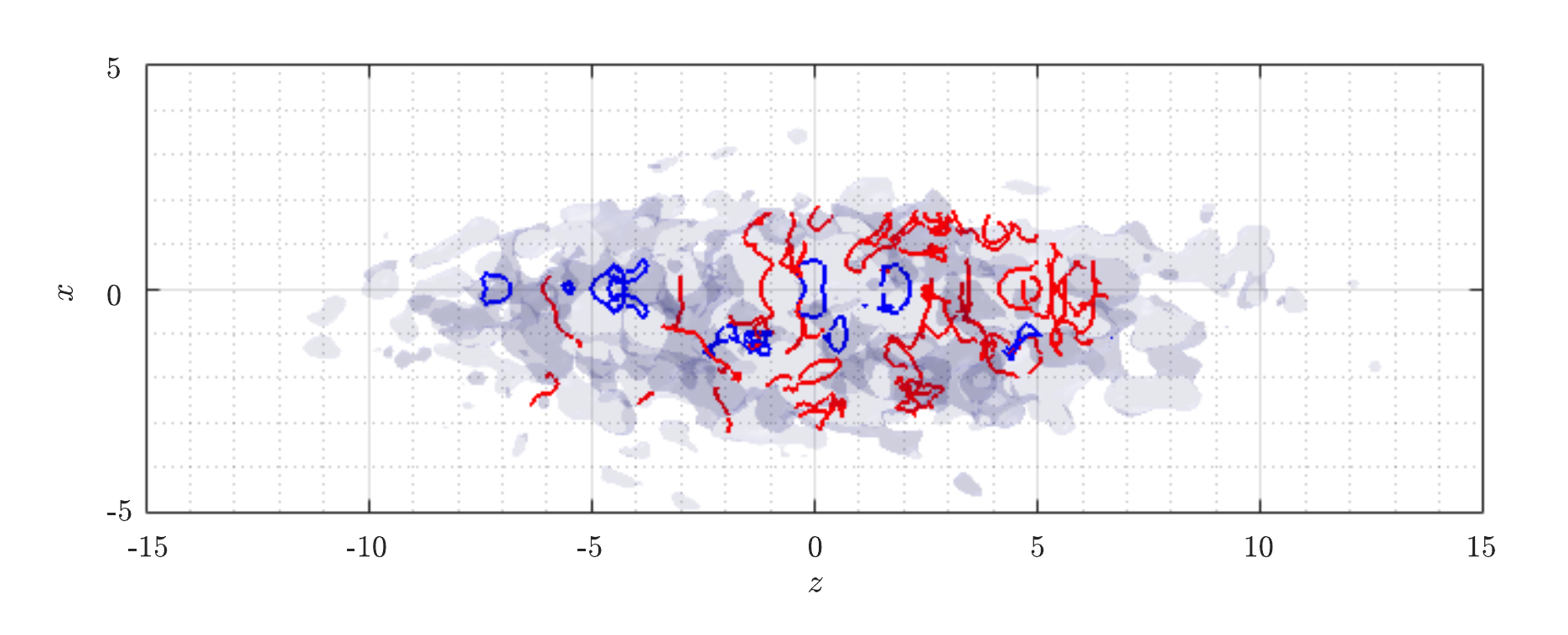}}
\caption{{The isosurface of the condensate at $t=35.2$ with the central vortex lines marked. Line vortices are denoted in red whereas ring vortices in blue. {Note that darker patches do not denote larger density, but reveal the fragmented nature of the condensate, as they result from the line of sight crossing the semi-transparent density isosurfaces multiple times.}}}
\label{fig:Fig5}
\end{figure*}

\begin{figure*}[]
\centering
{\includegraphics[trim={0.0cm 0.0cm 0.0cm 0cm},clip,width=0.9\textwidth]{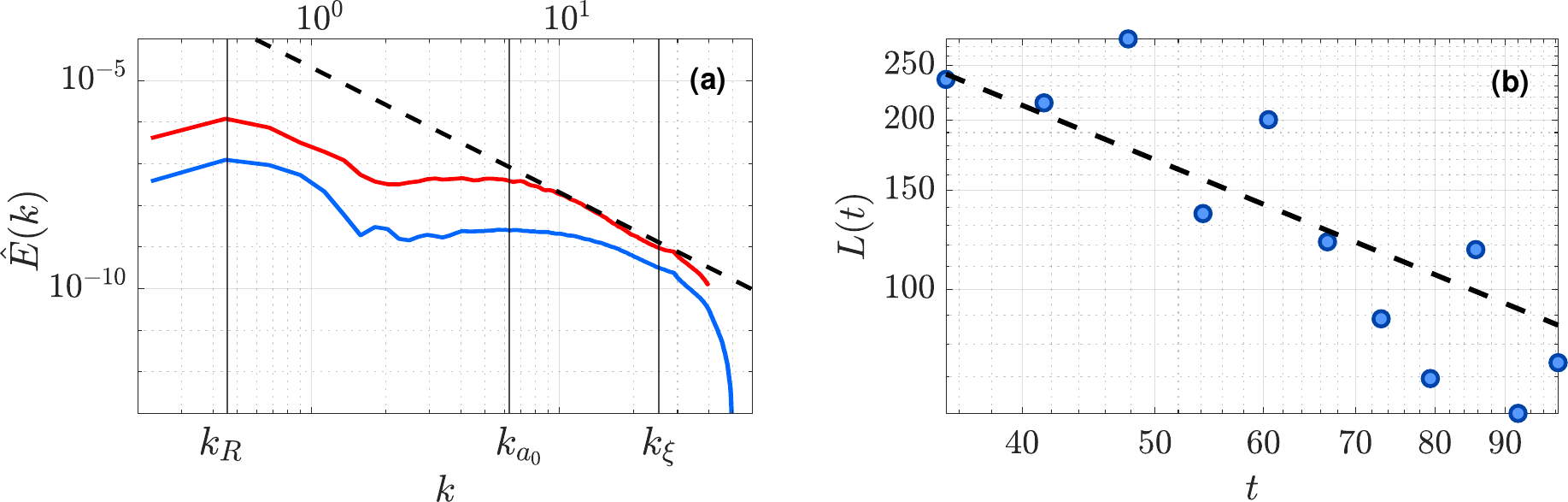}}
\caption{(a): numerical incompressible energy spectra $\hat{E}(k)$ of the condensate at times $t=35.2$ (red) and $t=92.0$ (blue). The dashed line, $\propto k^{-3}$, is drawn for reference. The spectra are shifted vertically for clarity. The vertical lines indicate the wavenumbers $k_{R}$, and $k_{\xi}$ corresponding in physical space to the longitudinal Thomas-Fermi radius $k_R=2\pi/11\approx 0.6$, and to the healing length $\xi$, respectively. The wavenumber $k_{a_0}$ corresponds to the vortex width. (b): total vortex line length $L(t)$ vs time $t$; the dashed line  $\propto t^{-1}$ is given for reference.}
\label{fig:Fig4}
\end{figure*}
\medskip
{\bf Vortex tangle.}
Besides large density fluctuations and fragmentation, the turbulent condensate contains vortex lines, clearly visible in the numerical simulations. Surprisingly, the vortex configuration is very different from turbulence in superfluid helium.
 
We identify the vortex lines in the simulation using a vortex tracking algorithm established elsewhere for homogeneous or smoothly-varying condensates \cite{rorai_skipper_kerr_sreenivasan_2016,villois2016vortex,PhysRevX.7.021031}. In our highly-fragmented system, it is numerically challenging to identify vortices when the condensate is off-centre in the trap (\textit{i.e.} when the condensate is most fragmented and loses a discernible shape). We hence focus on vortex reconstruction when the condensate lies at the centre of the trap.

The turbulent condensate and vortex lines therein at $t=35.2$ is shown in Fig. \ref{fig:Fig5}. It is apparent that there are two kinds of vortex lines: small vortex rings (shown as blue lines in the figure) and short open vortex lines which terminate at the condensate's boundary (shown as red lines). These open vortex lines are shorter versions of the {\it U-shaped vortices} discussed in the literature of non-turbulent condensates 
\cite{PhysRevLett.89.200403,modugno2003bose,PhysRevA.72.053624,PhysRevLett.115.170402}.
Both vortex rings and U-vortices are small, of the order of the vortex core size $a_0$, (see Supplementary Material (SM) A).

The lack of homogeneity of the vortex configuration is immediately visible in the figure. Although just after $T_D$  vortices are distributed more-or-less uniformly throughout the system, at later times most vortices reside at the back of the moving condensate (see SM B). When the longitudinal centre-of-mass, $\bar{z}$, is zero (Fig.~\ref{fig:Fig5}), the vortex rings tend to be located along the central $z$-axis of the condensate, while U-vortices tend to be more clustered towards the rear of the moving condensate which in Fig.~\ref{fig:Fig5} is moving towards left.

The orientation of the vortices, however, is fairly isotropic. The vortex length in each projected Cartesian direction falls between $30$ and $40\%$ of the total length at each time analysed. 
It is therefore fair to conclude that in a turbulent condensate the vortex tangle is isotropic but not homogeneous.

It is interesting to remark that, according to numerical simulations, small vortices have also been observed along the edges of a condensate excited by oscillating a boxtrap potential \cite{navon2016emergence}.
\bigskip

{\bf Energy spectrum.}
The current understanding of 3D quantum turbulence in superfluid helium arises from combined experimental, numerical and theoretical investigations  \cite{BarenghiLvovRoche} of the energy spectrum, $\hat{E}(k)$,
defined by $E_i=\int_0^{\infty} {\hat E}(k)dk$ where $k$ is the wavenumber
and $E_i$ is the total, incompressible, turbulent kinetic energy. The importance of ${\hat E}(k)$ is that it describes the energy distribution over the length scales, thus revealing inter-scale energy transfers. The key property of classical turbulence (described by the incompressible Navier-Stokes equation) is the celebrated Kolmogorov scaling
${\hat E}(k)\sim k^{-5/3}$.
In superfluid helium there appear to be two limiting regimes of quantum turbulence \cite{BarenghiSergeevBaggaley}: a {\it Kolmogorov regime} characterized by the same ${\hat E}(k) \sim k^{-5/3}$ observed in classical turbulence, indicating the existence of an energy cascade; and a {\it Vinen regime}, which is akin to a random flow, in which the energy spectrum peaks at the mesoscales $\Delta$ and decays as $k^{-1}$ at lengthscales smaller than $\Delta$ and larger than the vortex core $a_0$.

Unfortunately, the energy spectrum of turbulent 3D condensates is experimentally unavailable due to the lack of local velocity probes. There are also two significant differences with respect to liquid helium.
Firstly, condensates are very compressible and become easily fragmented (whereas the Kolmogorov scaling assumes constant density). Secondly, condensates are relatively small, so the spectrum extends only over a limited range of length scales, hindering any scaling law. In this respect, the comparison with liquid helium is staggering: the spatial extension of an atomic condensate is typically of the order of $10^2$ times the size of a vortex core, whereas in helium experiments \cite{Rousset2014} the size of the system can be as high as $10^{10}$ vortex cores. 

To make progress, in our numerical part of the study, we obtain the energy spectrum $\hat{E}(k)$ of the turbulent condensate by extracting the incompressible kinetic energy contribution, $E_i$, from the total kinetic energy via a standard Helmholtz decomposition \cite{nore1997decaying}.
The spectrum computed at two different times $t>T_D$ when $\bar{z}=0$, is reported in Fig. \ref{fig:Fig4}(a), showing no significant temporal dependence. The wavenumber corresponding to the average radius of the vortices at $t=35.2$, defined as $L/2\pi$ for vortex rings and $L/\pi$ for U-shaped vortices is very close to $k_{a_0}$. In the range $1.5 k_{a_0} \approx 10 < k < k_{\xi} \approx 25$ the energy spectrum scales approximately as $k^{-3}$, while no other scaling is observed at larger scales. At later times ($t=92$) the average radius decreases with the range of the $k^{-3}$ scaling decreasing accordingly. The $k^{-3}$ scaling, in contrast to both Kolmogorov's and Vinen's spectra, directly stems from the small size of the vortices observed in our turbulent condensate. In fact, the $k^{-3}$ spectrum reported between $k_{a_0}$ and $k_\xi$ coincides, as expected, with the incompressible kinetic energy spectrum inside the core of a quantum vortex \cite{nore1997decaying}. On the other hand, at smaller $k$, we lack the $k^{-1}$ spectrum (which one would assume given the random orientation of the vortices \cite{PhysRevA.96.023617}) precisely because the radii of the vortex rings are of the order of vortex core: there is no separation of scales between the radii of the vortex rings and their core, essential in order to observe the $k^{-1}$ scaling, (see SM C). Indeed, if we compute the spectrum of a homogeneous gas of small vortex rings, we recover the same $k^{-3}$ scaling for $k_{a_0} \lesssim k \lesssim k_{\xi}$, without further scalings at large scales (see SM C).

\bigskip

 {\bf Vortex decay.}
The random character of the vortex tangle is confirmed by the computed temporal decay of the total vortex length, which is itself a characteristic feature of the turbulent state. Indeed, by measuring the temporal behaviour of the vortex length at all times when $\bar{z}=0$, we find $L(t) \sim t^{-1 \pm 0.2}$ (Fig. \ref{fig:Fig4}(b)) matching the length temporal decay in the Vinen (random flow) regime, observed both in helium \cite{walmsley2008quantum} and numerically in atomic condensates when the turbulence is created by the instability of antiparallel multicharged vortices \cite{PhysRevA.96.023617}. This temporal decay behaviour is clearly distinct from the $L(t) \sim t^{-3/2}$ decay observed in Kolmogorov superfluid helium turbulence \cite{PhysRevLett.71.2583,PhysRevLett.82.4831,vinen2002quantum}.

{\bf Discussion.} The regime of 3D quantum turbulence characterising ongoing experiments on Bose-Einstein condensates excited via large-scale forcing \cite{navon2016emergence,PhysRevLett.103.045301}  raises theoretical challenges. 
The turbulence which we have identified in our experimental and numerical study in fact reveals the presence of large amplitude density waves, fragmentation and a vortex tangle composed of very small, randomly oriented vortex loops and lines, non-homogeneously distributed throughout the condensate. The small size of these vortical structures produces an incompressible energy spectrum exhibiting a $k^{-3}$ scaling at small scales, stemming from the properties of quantum vortex cores \cite{nore1997decaying}, and lacking any additional scaling at large lengthscales, given the striking absence of long vortex lines (or even bundles of lines) which have been observed in experiments and numerical simulations of quantum turbulence in superfluid helium. 

Overall, quantum turbulence generated in current experiments performed on confined condensates clearly appears to be distinct from the traditional turbulent regimes identified in other systems, \textit{i.e.}
Kolmogorov turbulence in both classical incompressible viscous
fluids \cite{Frisch1995} and superfluid helium \cite{BarenghiLvovRoche}, Vinen turbulence in superfluid helium \cite{walmsley2008quantum},
weak wave turbulence \cite{Nazarenko2015}
in classical fluids (ocean waves, acoustic waves, etc) and
in quantum fluids (Kelvin waves on  vortex lines \cite{LvovNazarenko2010,Krstulovic2012}). None of these turbulence
classes does fully account for the turbulence observed in our study which remarkably displays a coexistence of several turbulent features.

In this perspective, the possibility of turbulence which combines vortex lines and weakly nonlinear density waves was indeed suggested years ago \cite{NazarenkoOnorato2006} for an idealized 2D homogeneous system. However, what we have found in an actual
3D turbulent atomic condensate is a more radical combination of unusually strong density waves and unusually small vortex loops (`unusual' in the sense of previous paradigms) which
creates a scenario not seen in previous turbulence studies.
This creates a window of opportunities to search for more
possible phenomena that can only be present when different behaviors coexist. This the important conclusion of this study.

Future investigations should concentrate on the vortex nucleation mechanism, on how the turbulent features depend on the driving mechanism, and on the effect of thermal atoms. 
\bigskip

\centerline{\bf Methods}
\medskip

\centerline{\bf 1. Experiments.}
\bigskip

\noindent
A cloud of $N=3\times 10^5$ $^{87}$Rb atoms is confined in a magneto-optical trap, MOT1, cooled to around 140$\mu$K, and then transferred by radiation to a second magneto-optical trap, MOT2, which captures, accumulates and allows cooling to temperatures of a few $\mu$K. After this new entrapment and cooling, the optical fields are switched off, and the atoms are transferred to a magnetic trap composed of several coils forming a so-called IOFFE-PRITCHARD trap. Once in this trap, RF fields operating at a few MHz promote transitions causing the sample to evaporate. The loss of atoms during this phase is compensated by cooling, reaching temperatures of the order of 100nK, where condensate begins and progresses, until we have a condensate fraction ranging from 50\% to 80\% of the total atoms. The condensate is trapped into a elongated harmonic potential with $\omega_z=21\times2\pi$ rad/s and $\omega_r=130\times2\pi$ rad/s.

Once the condensate has been successfully cooled within the final trap, a secondary oscillating magnetic field with $\omega_{exc}=132.8\times2\pi$ rad/s is applied to the condensate. Here, a pair of anti-Helmholtz coils is applied close to the longitudinal axes of the static trap. The condensate is driven for a time of {37.65ms}, before the anti-Helmholtz coils are turned off and the system is left to evolve in the IOFFE-PRITCHARD trap. After $T_H$, the sample is allowed to expand freely for a period of 30ms. At the end of this free-flight, a resonant probe laser takes an absorption image, revealing the 2D projection of the expanded density. This projection allows the extraction of the momentum distribution as well as the fluctuation profile. For more details on the measurements and experimental techniques we refer to Methods and Refs. \cite{PhysRevA.79.043618,PhysRevLett.103.045301,Seman2011RouteTT}).
\newline
\centerline{\bf 2. Numerical simulations.}
\medskip
\noindent
\begin{figure}
    \centering
{\includegraphics[trim={0cm 0cm 0cm 0cm},clip,width=0.5\textwidth]{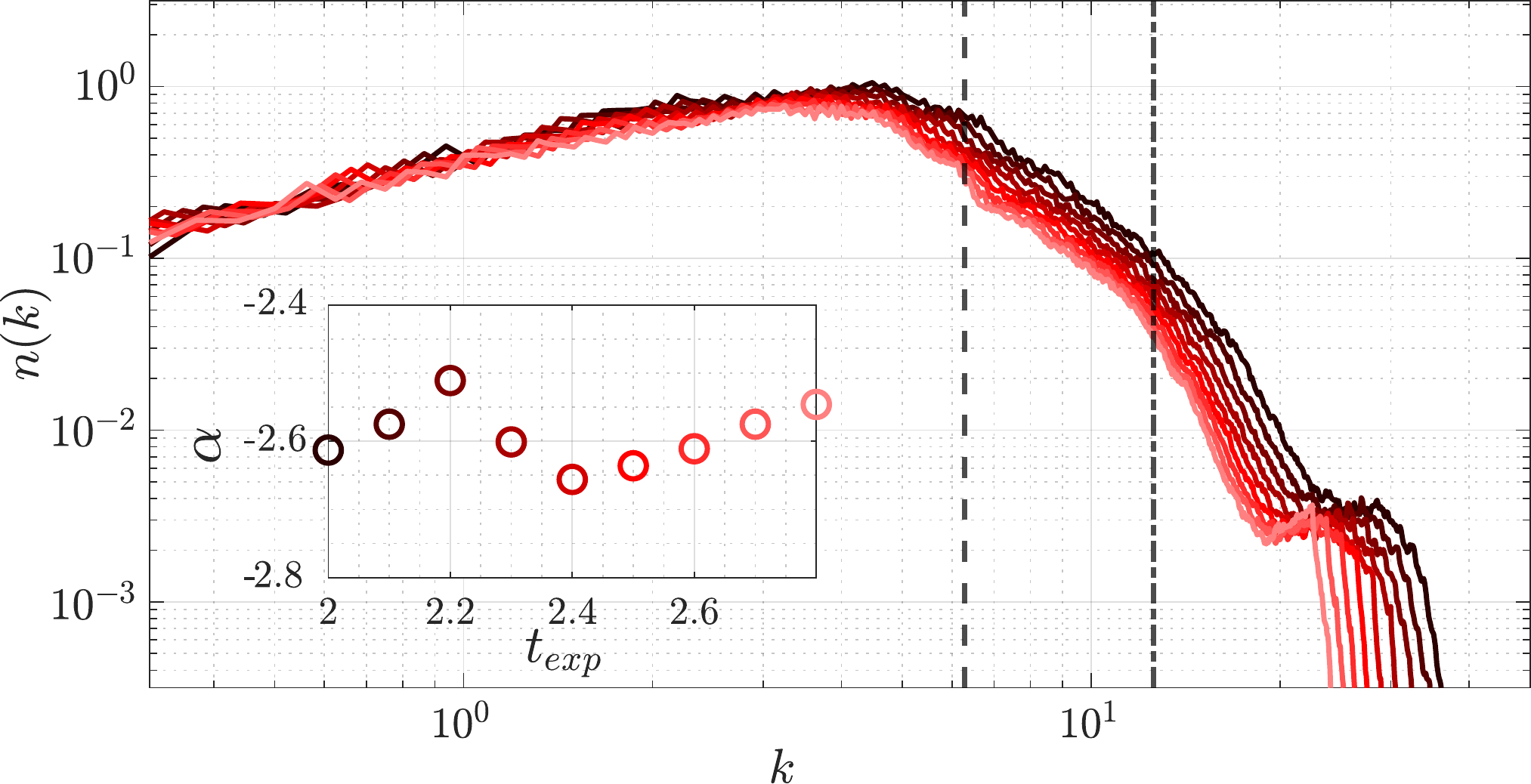}}
    \caption{{The momentum distributions from $t_{exp}=2.0$ to $2.8$ ($t=68.0$ to $68.8$) (going from dark to light) with the boundaries of fitting marked by $k=2\pi/4\xi$ (vertical dashed line) and $k=2\pi/2\xi$ (vertical dot-dashed line) and (inset) the power exponents for each of the momentum spectra in the corresponding colour.}}
    \label{fig:alpha}
\end{figure}

At temperatures much lower than the critical temperature, an atomic condensate of $N$ atoms with atomic mass $m$ and scattering length $a$ is quantitatively described by the Gross-Pitaevskii equation (GPE). It is convenient to rewrite the GPE in a dimensionless form using $\ell=(\hbar/(m \omega_r))^{1/2}$ as the unit of length, $1/\omega_r$ as the unit of time, and $N/\ell^3$ as the unit of the density $n=\vert \Psi \vert^2$, obtaining the dimensionless Gross-Pitaevskii equation (GPE) (\ref{eq:gpe_hou}) and the normalization $\int|\Psi|^2d\mathbf{x}=1$.
\newline
{The parameters are chosen to match those of the experiment; the resulting dimensionless chemical potential and longitudinal trapping frequency are respectively $\mu=13$. and $\omega_z=0.16$. These values, combined with the large oscillation which is imposed to the condensate, would require a computational domain too large to simulate numerically. {For this reason we increase the longitudinal trapping frequency to $\omega_z=0.5$ and lower the chemical potential from $\mu=13$ to $\mu=8$}. The dimensionless
interaction parameter $\mathcal{C}$ is calculated assuming a Thomas-Fermi condensate so that $\mathcal{C}=\mu^{5/2}\frac{16\pi\sqrt{2}}{15}\frac{1}{\omega_z}=1715$. }
The GPE is solved using finite-differences on a 3D computational grid and fourth order Runge-Kutta time integration.
The mesh size and time step are $\Delta x=0.125$ (corresponding to half of the system's healing length at the centre of the trap) and $\Delta t=0.001$ respectively. To find the initial ground state of the system, we first run the GPE in imaginary time, before moving to real time to observe the dynamics of the system with a total simulation time of $t=T_D+T_H=100$. {The method of obtaining the spectral exponent is shown in Fig. \ref{fig:alpha}; the condensate is released at $t=66.0$, at which it expands outwards and the momentum distribution is obtained. The condensate expands for a time $t_{exp}$, and an exponent $\alpha$ is obtained from fitting a power law between $k=2\pi/4\xi$ and $k=2\pi/2\xi$. The exponent measured in this interval quickly ($t_{exp}\ge2.0$; $t\ge68.0$) converges. Plotted in Fig. \ref{fig:alpha} are the results of the momentum distribution from $t_{exp}=2.0$ to $2.8$ ($t=68.0$ to $t=68.8$) with the inset providing the individual fitted power exponents obtained from each respective distribution in this $k$-range. The reported power law in this paper is taken as the average of these nine exponents to give a result of $\alpha=-2.6\pm0.1$}.

{\bf Acknowledgements.}
H.A.J.M.-S thanks the Engineering and Physical Sciences Research Council of the UK (Grant No. EP/R51309X/1) for support. L.G, N.G.P, and C.F.B acknowledges the Engineering and Physical Sciences Research Council of the UK (Grant No. EP/R005192/1) for support. L.G. acknowledges the support of  Istituto Nazionale di Alta Matematica (Italy). A.D.G.O, M.M., L.A.M, and V.S.B. acknowledges financial support from FAPESP (Grant 2013/07276-1) and CNPq - Brazilian Agencies. This research made use of the Rocket High Performance Computing service at Newcastle University.
\bibliography{apssamp}

\end{document}